\documentclass[12pt,letterpaper]{article}
\usepackage[dvips]{graphicx,color}
\input{psfig}
\input{epsf}
\usepackage{amsmath,amssymb}
\usepackage[dvips,letterpaper,text={6.5in,9in}]{geometry}
\usepackage{fancyhdr}
\usepackage{verbatim}

\newcommand\ltap{\
  \raise.3ex\hbox{$<$\kern-.75em\lower1ex\hbox{$\sim$}}\ } 
\newcommand\gtap{\
  \raise.3ex\hbox{$>$\kern-.75em\lower1ex\hbox{$\sim$}}\ } 

\newcommand\simge{\mathrel{%
   \rlap{\raise 0.511ex \hbox{$>$}}{\lower 0.511ex \hbox{$\sim$}}}}
\newcommand\simle{\mathrel{
   \rlap{\raise 0.511ex \hbox{$<$}}{\lower 0.511ex \hbox{$\sim$}}}}

\newcommand{\slashchar}[1]%
        {\kern .25em\raise.18ex\hbox{$/$}\kern-.75em #1}
\def\lsim{\mathrel{\raise.3ex\hbox{$<$\kern-.75em\lower1ex\hbox{$\sim$}}}}
\def\gsim{\mathrel{\raise.3ex\hbox{$>$\kern-.75em\lower1ex\hbox{$\sim$}}}}

\newcommand\CC{{\cal C}}

\newcommand\CH{{\cal H}}

\newcommand\CM{{\cal M}}

\newcommand\be{\begin{equation}} 
\newcommand\ee{\end{equation}} 
\newcommand\bea{\begin{eqnarray}}
\newcommand\eea{\end{eqnarray}}
\newcommand\ba{\begin{array}}
\newcommand\ea{\end{array}}
\newcommand\nn{\nonumber}

\newcommand{\half}{\ensuremath{\frac{1}{2}}}

\newcommand\ts{\thinspace}
\newcommand\ra{\rightarrow}

\newcommand\ol{\bar}

\newcommand\gev{{\rm GeV}}
\newcommand\tev{{\rm TeV}}

\newcommand\pb{{\rm pb}}
\newcommand\ipb{{\rm pb}^{-1}}

\newcommand\ifb{{\rm fb}^{-1}}
\newcommand\ecm{\sqrt{s}}
\newcommand\rshat{\sqrt{\shat}}
\newcommand\shat{\hat s}

\newcommand\suc{SU(3)_C}
\newcommand\Ntc{N_{TC}}
\newcommand\sutc{SU(N_{TC})}

\newcommand\atc{\alpha_{TC}}

\newcommand\Metc{M_{ETC}}

\newcommand\Ltc{\Lambda_{TC}}

\newcommand\tom{\omega_{T}}
\newcommand\tro{\rho_{T}}
\newcommand\atro{\alpha_{\tro}}

\newcommand\tropm{\rho_{T}^\pm}

\newcommand\troz{\rho_{T}^0}
\newcommand\tpi{\pi_T}
\newcommand\tpipm{\pi_T^\pm}
\newcommand\tpimp{\pi_T^\mp}
\newcommand\tpip{\pi_T^+}
\newcommand\tpim{\pi_T^-}
\newcommand\tpiz{\pi_T^0}
\newcommand\tpipr{\pi_T^{0 \prime}}

\newcommand\chipr{\chi^{\ts \prime}}

\newcommand\jet{{\rm jet}}

\begin{document}
\title{
\vskip -15mm
\begin{flushright}
\vskip -15mm
{\small BUHEP-06-01\\
hep-ph/0605119\\}
\vskip 5mm
\end{flushright}
{\Large{\bf Search for Low-Scale Technicolor at the Tevatron}}\\
}
\author{
{\large Kenneth Lane\thanks{lane@bu.edu}}\\
{\large Department of Physics, Boston University}\\
{\large 590 Commonwealth Avenue, Boston, MA 02215}\\
}
\maketitle
\begin{abstract}
  CDF and D\O\ each have more than $1\,\ifb$ of data on tape, and their
  stores are increasing. This should be sufficient to carry out significant
  searches for low-scale technicolor in $\tro \ra W \tpi$ and $\tom\,,\tro
  \ra \gamma \tpi$, processes whose cross sections may be as large as several
  picobarns. In this note we motivate and describe the Technicolor Straw Man
  framework for these processes and we urge that they be sought soon in the
  Run~2 data.\footnote{This paper is a contribution to the TeV4LHC Landscapes
    project.}

\end{abstract}


\newpage

\section*{1. Preamble}

\begin{figure}[t]
  \vspace{9.0cm}
  \includegraphics{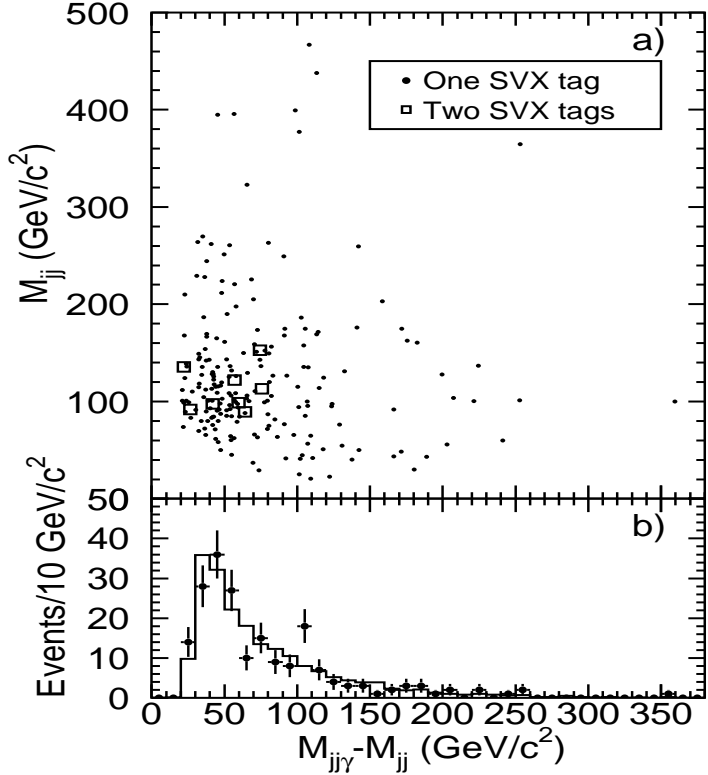}
  \vskip2.0truecm
  \caption{(a) The distribution of $M_{jj}$ vs. $M_{jj\gamma} - M_{jj}$ for events
    with a photon, $b$--tagged jet and a second jet. (b) Projection of this
    data in $M_{jj\gamma} - M_{jj}$; from Ref.~\cite{Abe:1998jc}.
    \label{TeV4LHC_fig_1} }
\end{figure}
 \begin{figure}[!ht]
   \begin{center}
     \includegraphics[width=3.20in, height = 3.20in, angle=0]
     {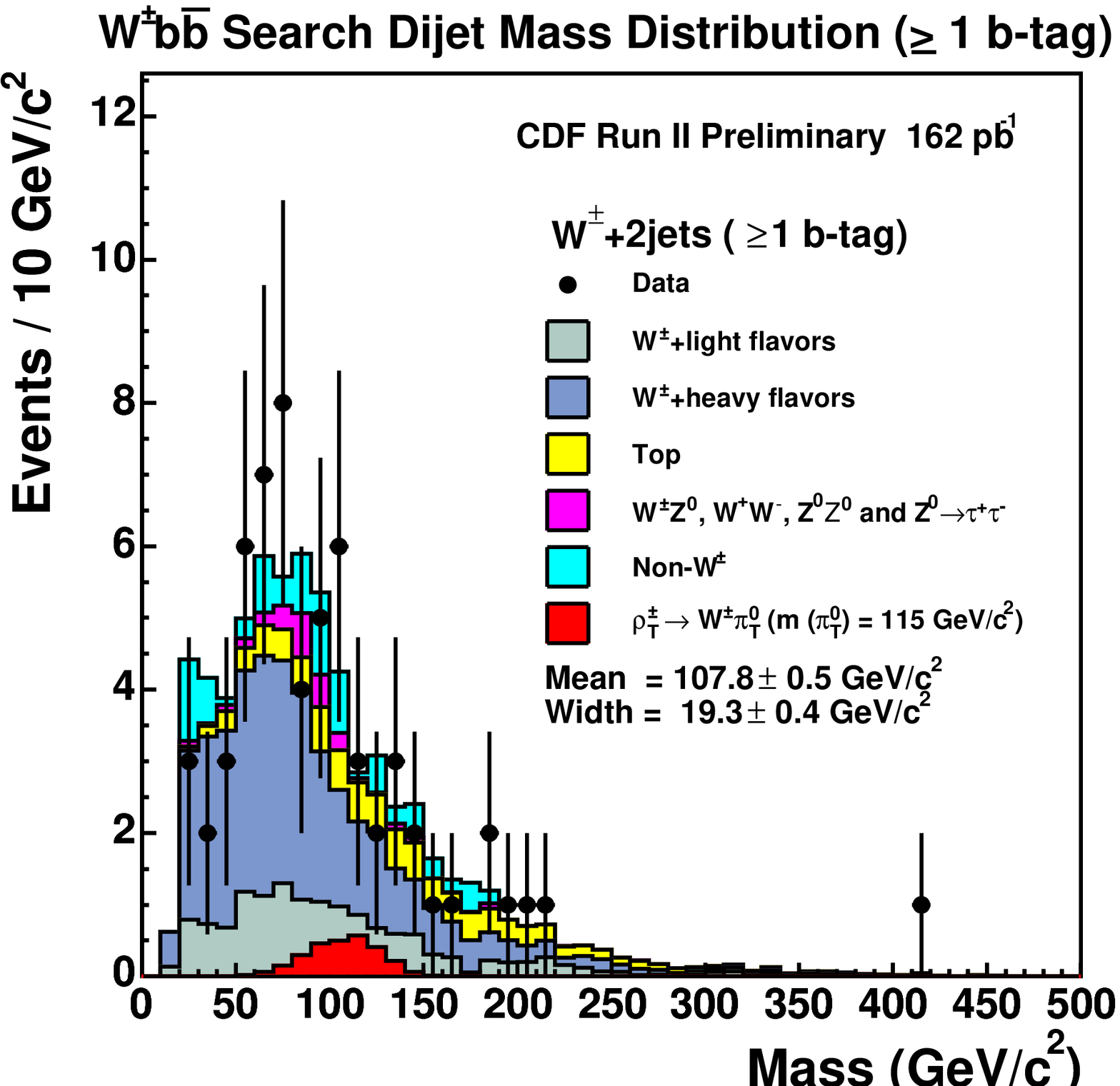}
     \includegraphics[width=3.20in, height = 3.20in, angle=0]
     {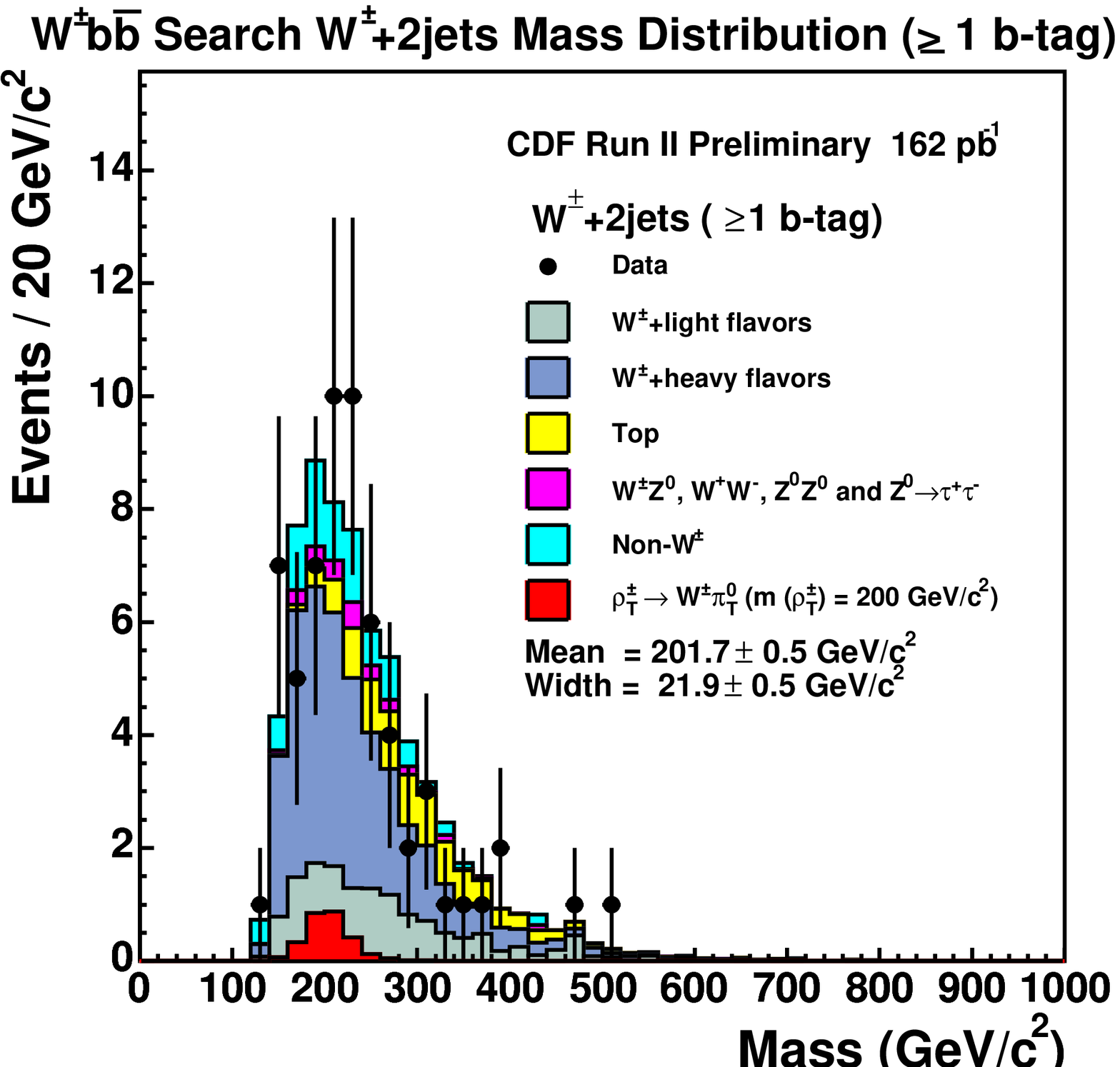}
     \caption{(a) Invariant mass of the dijet system with $\ge 1$ $b$-tagged
       jets; (b) Invariant mass of the $W+2\ts\jet$ system for the
       $\ell+2\ts\jet$ mode with $\ge 1$ $b$-tagged jets. From Run~2 with
       $162\,\ipb$; see {\tt
         http://www-cdf.fnal.gov/physics/exotic/r2a/20040722.lmetbj-wh-tc/} }
 \label{fig:TeV4LHC_fig_2}
   \end{center}
 \end{figure}
 
 Take a look at Figs.~\ref{TeV4LHC_fig_1} and~\ref{fig:TeV4LHC_fig_2}.
 Figure~\ref{TeV4LHC_fig_1} is from CDF in Run~1. It shows a search for $\tom
 \ra \gamma \tpi$, with $\tpi \ra b+$jet, based on about $100\,\ipb$,
 published in 1999~\cite{Abe:1998jc}. Note the $\sim 2\sigma$ excess near
 $M_{jj\gamma} - M_{jj} = 100\,\gev$. This search has not been repeated in
 Run~2.\footnote{Both detectors induce jet backgrounds to photons that
   require much effort to suppress. I hope that effort will be made.}
 Figure~\ref{fig:TeV4LHC_fig_2} is from CDF in Run~2. It shows results of an
 unpublished CDF study looking for $\tro \ra W^\pm \tpi$.\footnote{CDF's
   Run~1 version of this search is published in Ref.~\cite{Affolder:1999du}}
 The data were posted in from July 2004 and are based on $162\,\ipb$. There
 are small excesses in the dijet and $Wjj$ masses near 110~GeV and 210~GeV,
 respectively. Assuming $M_{\tom} = M_{\tro} \simeq 230\,\gev$, and taking
 into account losses from semileptonic $b$-decays, the excesses in Figs.~1
 and ~2 are in about the right place for $M_{\tpi} \simeq 120\,\gev$.
 
 In December 2005, CDF search was reported for $WH$-production with $W\ra
 \ell \nu$ and $H\ra b \bar b$ (with a single $b$-tag), based on
 $320\,\ipb$~\cite{Abulencia:2005ep}. The dijet mass spectrum is shown in
 Fig.~\ref{fig:TeV4LHC_fig_3}.\footnote{I am grateful to Y.-K.~Kim and her
   CDF collaborators for providing this figure.} There is a $2\sigma$ excess
 at $M_{jj} \simeq 110\,\gev$. The $Wjj$ spectrum was not reported and is
 {\em still} not available. This is puzzling since it requires no additional
 analysis to do so. Never mind that the expected rate for a $\sim 100\,\gev$
 Higgs decaying to $b \bar b$ and produced in association with a $W$ is about
 0.1~pb. If the excess were real, it would correspond to a total $WH$ cross
 section of about 5~pb, about 50~times the expected cross section. {\em Note
   added:} A D\O\ search for $WH \ra \ell\nu b \bar b$ also shows an apparent
 excess at 110~GeV in the dijet mass distribution in which one jet is
 $b$-tagged; see Fig.~10 in Ref.~\cite{Sopczak:2006xw}.
 
 Now, a little excess here and a little excess there is nothing to write home
 about. But when the excesses all show up in the same place, it's time to
 check them out. Both experiments have collected almost $1.5\,\ifb$. This
 summer, CDF and D\O\ will present new results for searches for SUSY and
 other more recent fads --- ADD large extra dimensions, RS gravitons, little
 Higgs to name a few. They should present the searches for technicolor as
 well. The most likely processes and search modes are 
\bea\label{eq:TCsearch_modes}
&& \tro^\pm \ra W^\pm \tpi^0 \ra \ell^\pm \nu_\ell + b\bar b \\
&& \tro^0 \ra W^\pm \tpi^\mp \ra  \ell^\pm \nu_\ell + b\bar c\,,\,b\bar u\\
&& \tom\,,\tro^0 \ra \gamma \tpiz\,,\gamma\tpipr \ra \gamma b \bar b\\
&& \tom\,,\tro^0 \ra e^+ e^-\,,\,\mu^+\mu^-\,.
\eea
These processes (and more) are available in
{\textsc{Pythia}}~\cite{Sjostrand:2000wi,Sjostrand:2003wg}.

In the rest of this paper, I will motivate low-scale technicolor --- that
technihadrons may be much lighter than $\sim 1\,\tev$ and, in fact, may be
readily accessible at the Tevatron. Then I will describe the Technicolor
Straw Man Model (TCSM) and present some rate estimates for
the most important color-singlet processes. The TCSM is described in more
detail in Refs.~\cite{Lane:1999uh,Lane:2002sm}, and much of Sects.~3-4 is
lifted from the second of these.

 \begin{figure}[t]
   \begin{center}
     \includegraphics[width=3.20in, height = 3.20in,angle=0]
     {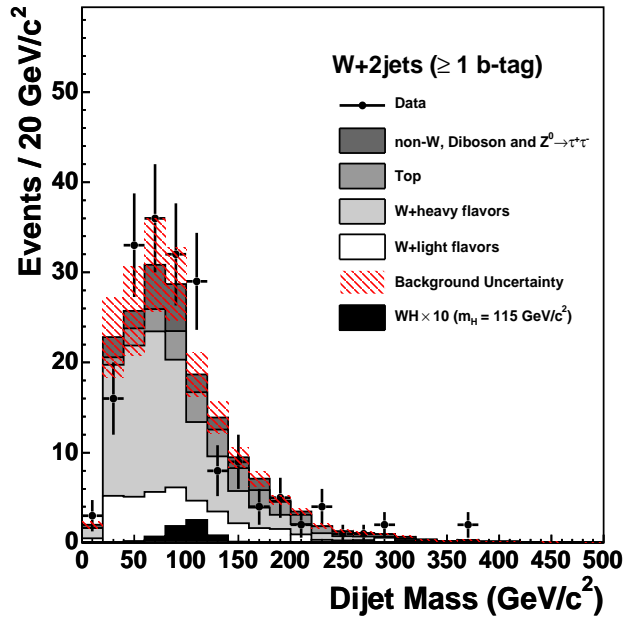}
     \caption{Invariant mass of the $W+2\ts\jet$ system for
      the $\ell+2\ts\jet$ mode with $\ge 1$ $b$-tagged jets; from Run~2 with
      $320\,\ipb$; see Ref.~\cite{Abulencia:2005ep}.}
     \label{fig:TeV4LHC_fig_3}
   \end{center}
 \end{figure}

\section*{2. Low-Scale Technicolor}

Technicolor is the theory of electroweak symmetry breaking (EWSB) by new
strong dynamics near 1~TeV~\cite{Weinberg:1979bn, Susskind:1978ms}. It is the
most natural scenario (not to mention the only one with a precedent, namely,
QCD) for dealing with the standard model's naturalness problem: it banishes
{\em elementary} scalar particles altogether. Technicolor by itself, however,
cannot explain --- or even describe in a phenomenological way, as the
standard model does --- the origin of quark and lepton masses and mixings.
The only known way to do that in the dynamical context of technicolor is
extended technicolor (ETC)~\cite{Eichten:1979ah}.

Two elements of the modern formulation of technicolor (see the reviews and
references in Refs.~\cite{Lane:2002wv, Hill:2002ap}) strongly suggest that
its energy scale $\Ltc \simeq 4\pi F_T$ --- and therefore the masses of
technihadrons ($\tro$ and $\tom$ as well as $\tpi$) --- are {\em much} less
than several TeV. They are the notions of {\em walking technicolor}
(WTC)~\cite{Holdom:1981rm,Appelquist:1986an, Yamawaki:1986zg,Akiba:1986rr}
and {\em topcolor-assisted technicolor} (TC2)~\cite{Hill:1995hp}. Here, $F_T$
is the technipion decay constant. Assuming for simplicity that the
technifermions form $N_D$ electroweak doublets, then $F_T \simeq
F_\pi/\sqrt{N_D}$, where $F_\pi = 246\,\gev$. The EWSB condensate is $\langle
\bar TT\rangle_{TC} \simeq 4\pi F_T^3$.

Extended technicolor inevitably induces flavor-changing neutral current
interactions of quarks and leptons. The most problematic of these are the
$|\Delta S| = 2$ operators,
\be\label{eq:dStwo}
\CH_{|\Delta S|=2} = \frac{g^2_{ETC}}{\Metc^2} \sum_{ij} K_{ij}\, \ol s
\Gamma_i d \,\ol s \Gamma_j d + {\rm h.c.} \,,
\ee
and they require effective ETC gauge boson masses $\Metc/g_{ETC}\sqrt{K_{ij}}
\simge 1000\,\tev$. If TC were a QCD-like gauge theory, one in which
asymptotic freedom sets in quickly near $\Ltc$, the quark and lepton masses
$m_{q,l} \simeq g^2_{ETC} \langle \ol TT\rangle_{ETC}/M^2_{ETC}$ generated by
such high-scale ETC interactions would be unacceptably small because $\langle
\bar TT\rangle_{ETC} \simeq \langle \bar TT\rangle_{TC}$. This difficulty is
cured by walking technicolor. In WTC, the technicolor gauge coupling $\atc$
runs very slowly, i.e., the interaction is close to conformally invariant,
and the technifermion condensates $\langle \ol TT\rangle_{ETC}$ renormalized
at the ETC scale are enhanced relative to $\langle \ol TT\rangle_{TC}$ by a
factor not much less than $\Metc/\Ltc$. The small $\beta_{TC}$-function
required for WTC is readily achieved by having many technidoublets
transforming as the fundamental representation of the TC gauge group. Then
$N_D$ is large and $F_T$ is small.\footnote{Walking could in principle be
  achieved by having a few technidoublets in higher-dimensional TC
  representations; see Refs.~\cite{Lane:1989ej} and~\cite{Dietrich:2005jn,
    Evans:2005pu}. It is difficult to see how this could be done without some
  number of doublets in the fundamental; see Ref.~\cite{Lane:1991qh}. Another
  option is to have a large number of technifermions in the fundamental TC
  representation, but only one doublet of them has electroweak
  interactions~\cite{Christensen:2005cb}.}

Even with the enhancements of walking technicolor, there is no satisfactory
way in the context of ETC alone to understand the large mass of the top
quark. Either the ETC mass scale generating $m_t$ must be too close to $\Ltc$
or the ETC coupling must be fine-tuned.\footnote{A possible exception to this
  has been proposed in Ref.~\cite{Appelquist:2003hn}. In this model, $N_D =
  4$ and $F_T$ is not particularly small. The model is genuinely baroque, but
  that is probably true of any quasi-realistic ETC model.} So far, the most
attractive scheme for $m_t$ is that it is produced by the condensation of top
quarks, induced at a scale near $1\,\tev$ by new strong topcolor gauge
interactions ($SU(3) \otimes U(1)$ in the simplest scheme). This top
condensation scheme, topcolor-assisted technicolor, accounts for almost all
the top mass, but for only a few percent of EWSB. Realistic models that
provide for the TC2 gauge symmetry breaking and for the mixing of the heavy
third generation with the two light generations typically require many ($N_D
\simeq 10$ (!)) technifermion doublets. Therefore, in the following, we shall
assume $F_T \simle 100\,\gev$.\footnote{The question of the effect of
  technicolor on precisely measured electroweak quantities such as $S$, $T$,
  and $U$ naturally arises because of the appearance of many technifermion
  doublets in low-scale technicolor. Calculations that show technicolor to be
  in conflict with precision measurements have been based on the assumption
  that technicolor dynamics are just a scaled-up version of QCD. However,
  because of its walking gauge coupling, this cannot be. In walking
  technicolor there must be something like a tower of spin-one technihadrons
  reaching almost to the ETC scale, and these states must contribute
  significantly to the integrals over spectral functions involved in
  calculating $S$, $T$, and $U$. Therefore, in the absence of detailed
  experimental knowledge of this spectrum, including the spacing between
  states and their coupling to the electroweak currents, it has not yet been
  possible to estimate these quantities reliably.}

\section*{3. The Technicolor Straw Man Model}

The TCSM provides a simple framework for searching for light technihadrons.
Its first and probably most important assumption is that the lowest-lying
bound states of the lightest technifermions can be considered {\em in
  isolation}. The lightest technifermions are expected to be an isodoublet of
color singlets, $(T_U, T_D)$. Color triplets, not considered here, will be
heavier because of $\suc$ contributions to their hard (chiral symmetry
breaking) masses. We assume that all technifermions transform under
technicolor $SU(\Ntc)$ as fundamentals. This leads us to make --- with no
little trepidation in a {\em walking} gauge theory --- large-$\Ntc$ estimates
of certain parameters. The electric charges of $(T_U,T_D)$ are $Q_U$ and $Q_D
= Q_U - 1$; they are important parameters of the TCSM. The color-singlet
bound states we consider are vector and pseudoscalar mesons. The vectors
include a spin-one isotriplet $\tro^{\pm,0}$ and an isosinglet $\tom$.
Techni-isospin can be a good approximate symmetry in TC2, so that $\tro$ and
$\tom$ are nearly degenerate. Their mixing with each other and the photon and
$Z^0$ is described by a neutral-sector propagator matrix.

The lightest pseudoscalar bound states of $(T_U,T_D)$ are the color-singlet
technipions. They also form an isotriplet $\Pi_T^{\pm,0}$ and an isosinglet
$\Pi_T^{0 \prime}$. However, these are not mass eigenstates. Our second
important assumption for the TCSM is that the isovectors may be described as
simple {\em two-state mixtures} of the longitudinal weak bosons $W_L^\pm$,
$Z_L^0$ --- the true Goldstone bosons of dynamical electroweak symmetry
breaking --- and mass-eigenstate pseudo-Goldstone technipions $\tpi^\pm, \tpiz$:
\be\label{eq:pistates}
 \vert\Pi_T\rangle = \sin\chi \ts \vert
W_L\rangle + \cos\chi \ts \vert\tpi\rangle\ts.
\ee
We assume that $\sutc$ gauge interactions dominate the binding of all
technifermions into technihadrons. Then the decay constants of color-singlet
and nonsinglet $\tpi$ are approximately equal, $F_T \simeq F_\pi/\sqrt{N_D}$,
and the mixing factor $\sin\chi$ --- another important TCSM parameter --- is
given by
\be\label{eq:sinchi}
\sin\chi \simeq F_T/F_\pi \simeq 1/\sqrt{N_D}\ts,
\ee
so that $\sin^2\chi \ll 1.$

Similarly, $\vert\Pi_T^{0 \prime} \rangle = \cos\chipr \ts
\vert\tpipr\rangle\ + \cdots$, where $\chipr$ is another mixing angle and the
ellipsis refer to other technipions needed to eliminate the two-technigluon
anomaly from the $\Pi_T^{0 \prime}$ chiral current. It is unclear whether,
like $\tro$ and $\tom$, these neutral technipions will be degenerate.  If
$\tpiz$ and $\tpipr$ are nearly degenerate {\em and} if their widths are
roughly equal, there may be appreciable $\tpiz$--$\tpipr$ mixing and, then,
the lightest neutral technipions will be ideally-mixed $\ol T_U T_U$ and $\ol
T_D T_D$ bound states. {\em Searches for these technipions ought to consider
  both possibilities: they are nearly degenerate or that $M_{\pi_T^\pm} =
M_{\pi_T^0} \ll M_{\tpipr}$}.

Color-singlet technipion decays are mediated by ETC and (in the case of
$\tpipr$) $\suc$ interactions. In the TCSM they are taken to be:
\bea\label{eq:tpiwidths}
 \Gamma(\tpi \ra \ol f' f) &=& \frac{1}{16\pi F^2_T}
 \ts N_f \ts p_f \ts C^2_{1f} (m_f + m_{f'})^2 \nn \\ \nn \\
 \Gamma(\tpipr \ra gg) &=& \frac{1}{128 \pi^3 F^2_T}
 \ts \alpha_C^2 \ts C^2_{1g} \ts \Ntc^2 \ts M_{\tpipr}^{\frac{3}{2}} \ts .
\eea
Like elementary Higgs bosons, technipions are {\em expected} to couple to
fermion mass. Thus, $C_{1f}$ is an ETC-model dependent factor of order one
{\em except} that TC2 implies a weak coupling to top quarks, $\vert
C_{1t}\vert \simle m_b/m_t$. Thus, there is no strong preference for
technipions to decay to (or radiate from) top quarks. The number of colors of
fermion~$f$ is $N_f$. The fermion momentum is $p_f$. The QCD coupling
$\alpha_C$ is evaluated at $M_{\tpi}$; and $C^2_{1g}$ is a Clebsch of order
one. The default values of these and other parameters are tabulated in
Ref.~\cite{Lane:2002sm}. For $M_{\tpi} < m_t + m_b$, these technipions are
expected to decay mainly as follows: $\tpip \ra c \ol b$, $u \ol b$, $c \ol
s$ and possibly $\tau^+ \nu_\tau$; $\tpiz \ra b \ol b$ and, perhaps $c \ol
c$, $\tau^+\tau^-$; and $\tpipr \ra gg$, $b \ol b$, $c \ol c$,
$\tau^+\tau^-$.

In the limit that the electroweak couplings $g,g' = 0$, the $\tro$ and $\tom$
decay as
\bea\label{eq:vt_decays}
\tro &\ra& \Pi_T \Pi_T = \cos^2 \chi\ts (\tpi\tpi) + 2\sin\chi\ts\cos\chi
\ts (W_L\tpi) + \sin^2 \chi \ts (W_L W_L) \ts; \nn \\\nn\\
\tom &\ra& \Pi_T \Pi_T \Pi_T = \cos^3 \chi \ts (\tpi\tpi\tpi) + \cdots \ts.
\eea
The $\tro$ decay amplitude is
\be\label{eq:rhopipi}
\CM(\tro(q) \ra \pi_A(p_1) \pi_B(p_2)) = g_{\tro} \ts \CC_{AB}
\ts \epsilon(q)\cdot(p_1 - p_2) \ts,
\ee
where $\epsilon(q)$ is the $\tro$ polarization vector; $\atro \equiv
g_{\tro}^2/4\pi = 2.91(3/\Ntc)$ is scaled naively from QCD and the parameter
$\Ntc = 4$ is used in calculations; and
\be\label{eq:ccab}
\ba{ll}
\CC_{AB} &= \left\{\ba{ll} \sin^2\chi  & {\rm for} \ts\ts\ts\ts W_L^+ W_L^-
\ts\ts\ts\ts {\rm or} \ts\ts\ts\ts  W_L^\pm Z_L^0 \\
\sin\chi \cos\chi & {\rm for} \ts\ts\ts\ts W_L^\pm \tpimp\,,
\ts\ts\ts\ts  {\rm or} \ts\ts\ts\ts W_L^\pm \tpiz\,, Z_L^0 \tpipm \\
\cos^2\chi & {\rm for} \ts\ts\ts\ts \tpip\tpim  \ts\ts\ts\ts {\rm or}
\ts\ts \ts\ts \tpipm\tpiz \ts.
\ea \right.
\ea
\ee
The $\tro$ decay rate to two technipions is then (for use in cross
sections, we quote the energy-dependent width for a $\tro$ mass of
$\sqrt{\shat}$)
\be\label{eq:trhopipi}
\Gamma(\troz \ra \pi_A^+ \pi_B^-) = \Gamma(\tropm \ra \pi_A^\pm \pi_B^0) =
\frac{2 \atro \CC^2_{AB}}{3} \ts {\ts\ts \frac{p^3}{\shat}} \ts,
\ee
where $p = [(\shat - (M_A+M_B)^2) (\shat - (M_A-M_B)^2)]^\half/2\rshat$
is the $\tpi$ momentum in the $\tro$ rest frame.

\section*{4. Sample TCSM Production Rates at the Tevatron}

The $\tro \ra\Pi_T\Pi_T$ decays are strong transitions, and we might
therefore expect the $\tro$ to be quite wide. Almost certainly, this is not
so. The enhanced technifermion condensate in walking technicolor magnifies
technipions' masses much more than it does technivectors' and, so, the
channels $\tro \ra \tpi\tpi$, $\tom \ra \tpi\tpi\tpi$ and even the
isospin-violating $\tom \ra \tpi\tpi$ are likely to be
closed~\cite{Lane:1989ej}. A $\troz$ of mass $200\,\gev$ may then decay
mainly to $W_L^\pm \tpi^\mp$ or $W_L^+ W_L^-$. These channels are also
isospin-forbidden for the $\tom$, and so all its important decays are
electroweak: $\tom \ra \gamma\tpiz$, $Z^0\tpiz$, $W^\pm \tpimp$, and $\ol f
f$ --- especially $e^+e^-$ and $\mu^+\mu^-$. Here, the $Z$ and $W$ are
transversely polarized.\footnote{Strictly speaking, the identification of $W$
  and $Z$ decay products as longitudinal or transverse is approximate,
  becoming exact in the limit of very large $M_{\tro,\tom}$.} Furthermore,
since $\sin^2\chi \ll 1$, the electroweak decays of $\tro$ to the transverse
gauge bosons $\gamma,W,Z$ plus a technipion may be competitive with the
open-channel strong decays. Thus, we expect $\tro$ and $\tom$ to be
{\underbar{\em very narrow}}. For masses accessible at the Tevatron, it turns
out that $\Gamma(\tro) \sim 1\,\gev$ and $\Gamma(\tom)\simle 0.5\,\gev$.

Within the context of the TCSM (and with plausible assumptions for its
parameters), we expect that $\tro^{\pm,0}$ and $\tom$ with masses below about
$250\,\gev$ should be accessible in Tevatron Run~2 in one channel or another.
Assuming $M_{\rho_T} < 2M_{\pi_T}$, the $\tro \ra W\tpi$ cross sections have
rates of a few picobarns. An example is shown in Fig.~\ref{TeV4LHC_fig_4},
for $M_{\rho_T}= 210\,\gev$ and $M_{\pi_T} = 110\,\gev$.\footnote{This figure
  does not include contributions from transverse weak bosons, which are small
  for this choice of parameters.} The parameter $M_V$
against which these rates are plotted is described below; it hardly affects
them. These cross sections were computed with EHLQ structure
functions~\cite{Eichten:1984eu}, and they should be multiplied by a K-factor
of about~1.4, typical of Drell-Yan processes such as these. Searches for
these modes at the Tevatron require a leptonic decay of the $W$ plus two jets
with at least one $b$-tag.

\begin{figure}[t]
  \vspace{9.0cm}
  \includegraphics{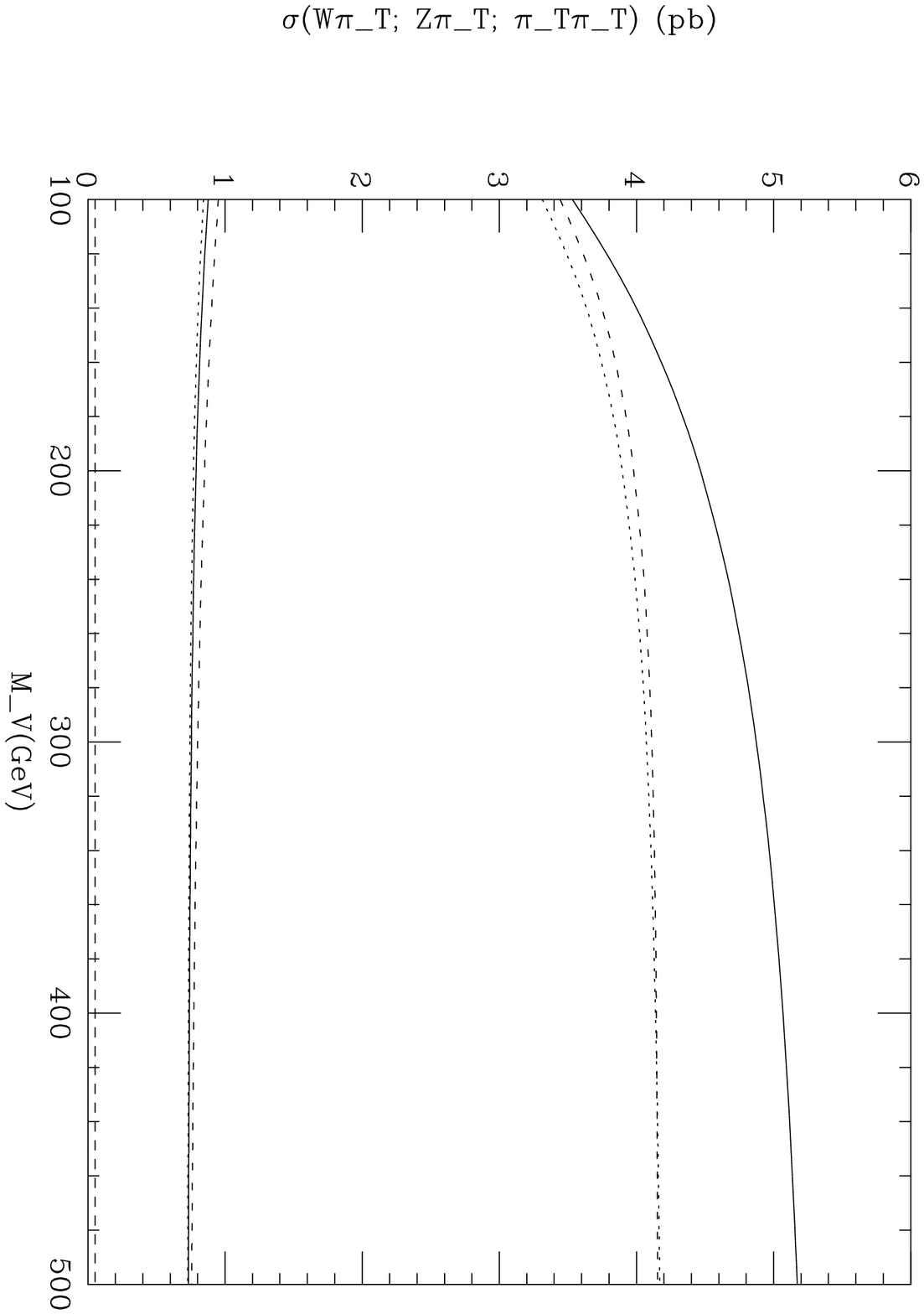}
\caption{Production rates rates in $p \ol p$ collisions at $\ecm = 2\,\tev$
  for $\tom$, $\troz$, $\tropm \ra W \tpi$ (upper
  curves) and $Z\tpi$ (lower curves) versus $M_V$, for $M_{\tro} = 210\,\gev$
  and $M_{\tom} = 200$ (dotted curve), 210 (solid), and $220\,\gev$
  (short-dashed); $Q_U + Q_D = \frac{5}{3}$ and $M_{\tpi} = 110\,\gev$. Also shown
  is $\sigma(\tro \ra \tpi\tpi)$ (lowest dashed curve); from Ref.~\cite{Lane:1999uh}.
  \label{TeV4LHC_fig_4}}
\end{figure}

\begin{figure}[!ht]
  \vspace{9.0cm}
  \includegraphics{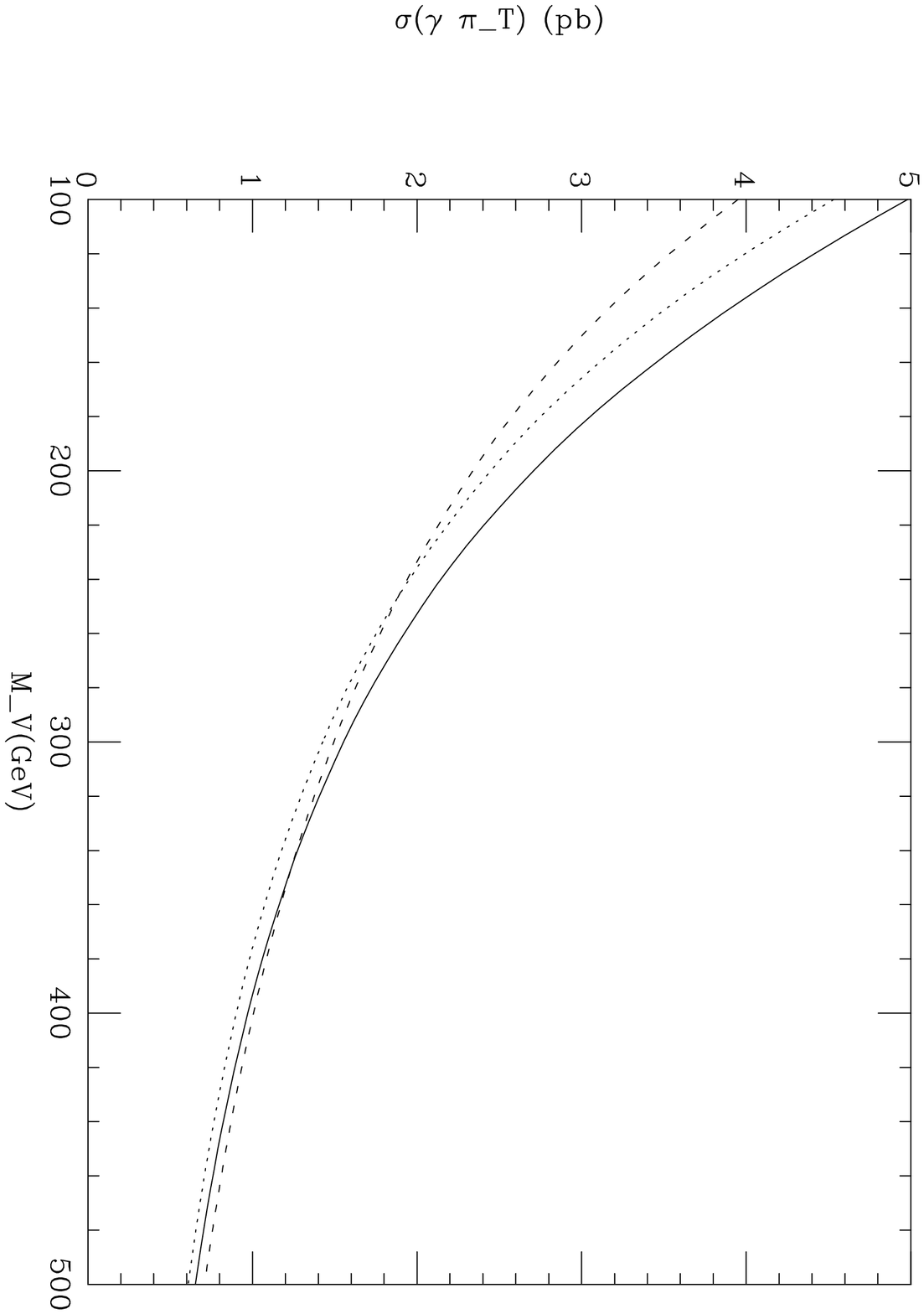}
  \caption{Production rates in $p \ol p$ collisions at $\ecm = 2\,\tev$
for the sum of $\tom$, $\troz$, $\tropm \ra \gamma
  \tpi$ and $\gamma\tpipr$ versus $M_V$, for $M_{\tro} = 210\,\gev$ and
  $M_{\tom} = 200$ (dotted curve), 210 (solid), and $220\,\gev$
  (short-dashed); $Q_U + Q_D = \frac{5}{3}$, and $M_{\tpi} = M_{\tpipr} = 110\,\gev$;
  from Ref.~\cite{Lane:1999uh}.
  \label{TeV4LHC_fig_5}}
\end{figure}

The parameter $M_V$ appears inversely in the amplitude for $\tro,\tom \ra
\gamma\tpi$. It is a typical TC mass-scale and, for low-scale technicolor,
should lie in the range 100--$500\,\gev$. As long as the $\tro\ra W\tpi$
channels are open, $\gamma\tpiz$ and $\gamma\tpipr$ production proceeds mainly
through the $\tom$ resonance. Then $M_V$ and the sum of the technifermion
charges, $Q_U + Q_D$, control their rates, which are approximately
proportional to $(Q_U+Q_D)^2/M_V^2$. Figure~\ref{TeV4LHC_fig_5} shows the
$\gamma\tpi$ cross sections vs.~$M_V$ for the favorable case $Q_U + Q_D =
\frac{5}{3}$. Again, a K-factor of about 1.4 should be applied. Here,
$M_{\tpipr} = M_{\tpiz}$ and about half the rate is $\gamma\tpipr$. Note that the
$gg$ decays of of the $\tpipr$ will dilute the usefulness of the $b$-tag for
these processes. On the other hand, decays involving $b$'s have two $b$-jets.

Finally, for large $M_V$, $\tom$ decays mainly to $\bar ff$ pairs. The most
promising modes at the Tevatron (and the LHC) then are $e^+e^-$ and
$\mu^+\mu^-$. Figures~\ref{TeV4LHC_fig_6} and~\ref{TeV4LHC_fig_7} show the
effect of changing $M_V$ from~100 to $500\,\gev$ on the $e^+e^-$ invariant
mass distributions. Note also the $\tom$--$\tro$ interference effect when
their masses are close. This would be lovely to observe! The cross section,
for $M_{\tom} = M_{\tro} = 210\,\gev$, integrated from 200 to $220\,\gev$,
and including the Drell-Yan background, increases from 0.12 to $0.25\,\pb$
when $M_V$ is increased from 100 to $500\,\gev$. A first search was for
$\tom,\tro\ra e^+e^-$ was carried out by D\O\ in Run~1 and published in
Ref.~\cite{Abazov:2001qd}. I look forward to a search based on Run~2 data
soon. It shouldn't be difficult to carry out.

\begin{figure}[!ht]
  \vspace{9.0cm}
  \includegraphics{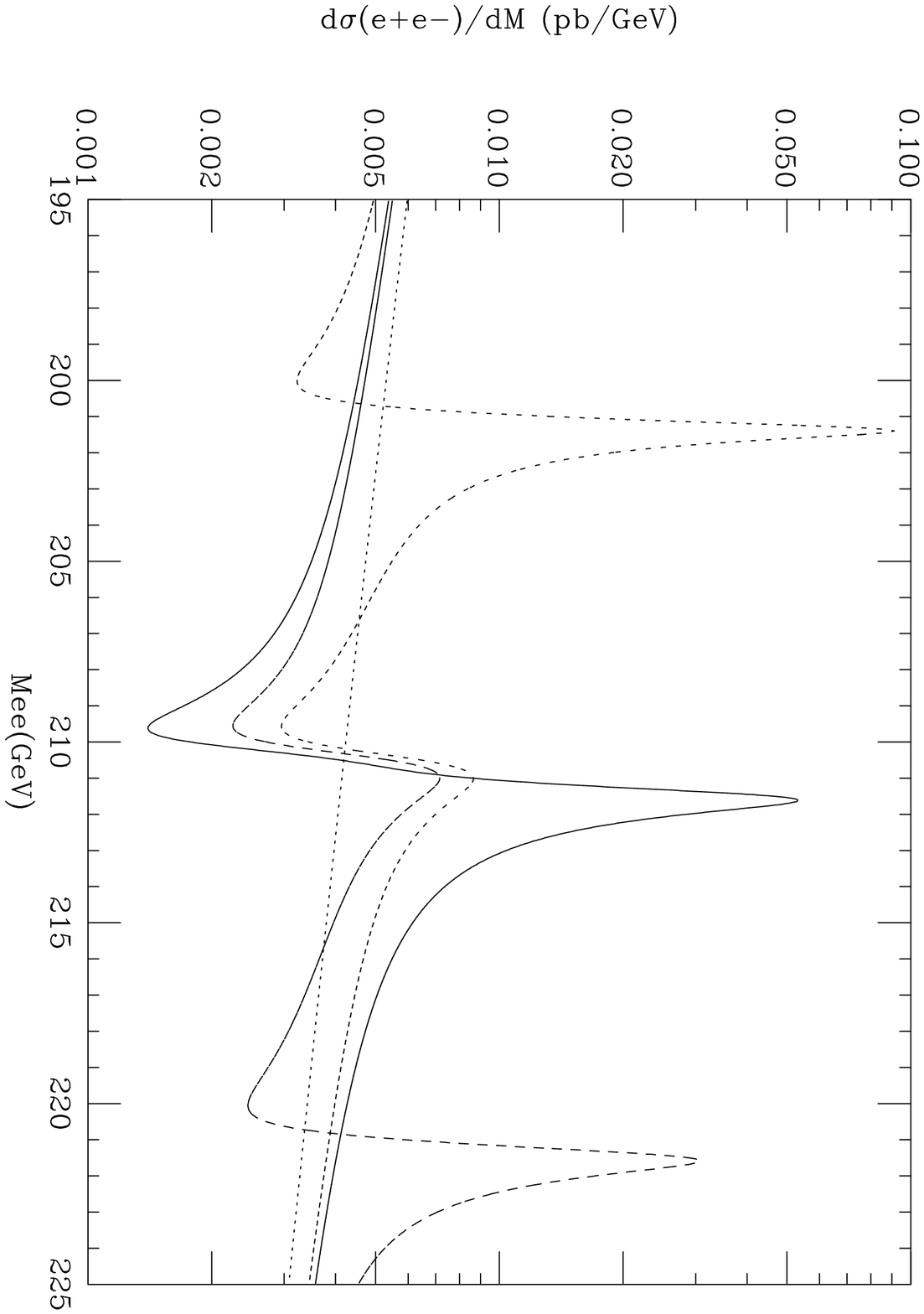}
\caption{Invariant mass distributions in $p \ol p$ collisions at $\ecm =
  2\,\tev$ for $\tom$, $\troz \ra e^+e^-$
  for $M_{\tro} = 210\,\gev$ and $M_{\tom} = 200$ (short-dashed curve), 210
  (solid), and $220\,\gev$ (long-dashed); $M_V = 100\,\gev$. The standard
  model background is the sloping dotted line. $Q_U + Q_D = \frac{5}{3}$ and
  $M_{\tpi} = 110\,\gev$; from Ref.~\cite{Lane:1999uh}.
  \label{TeV4LHC_fig_6}}
\end{figure}

\begin{figure}[!ht]
  \vspace{9.0cm}
  \includegraphics{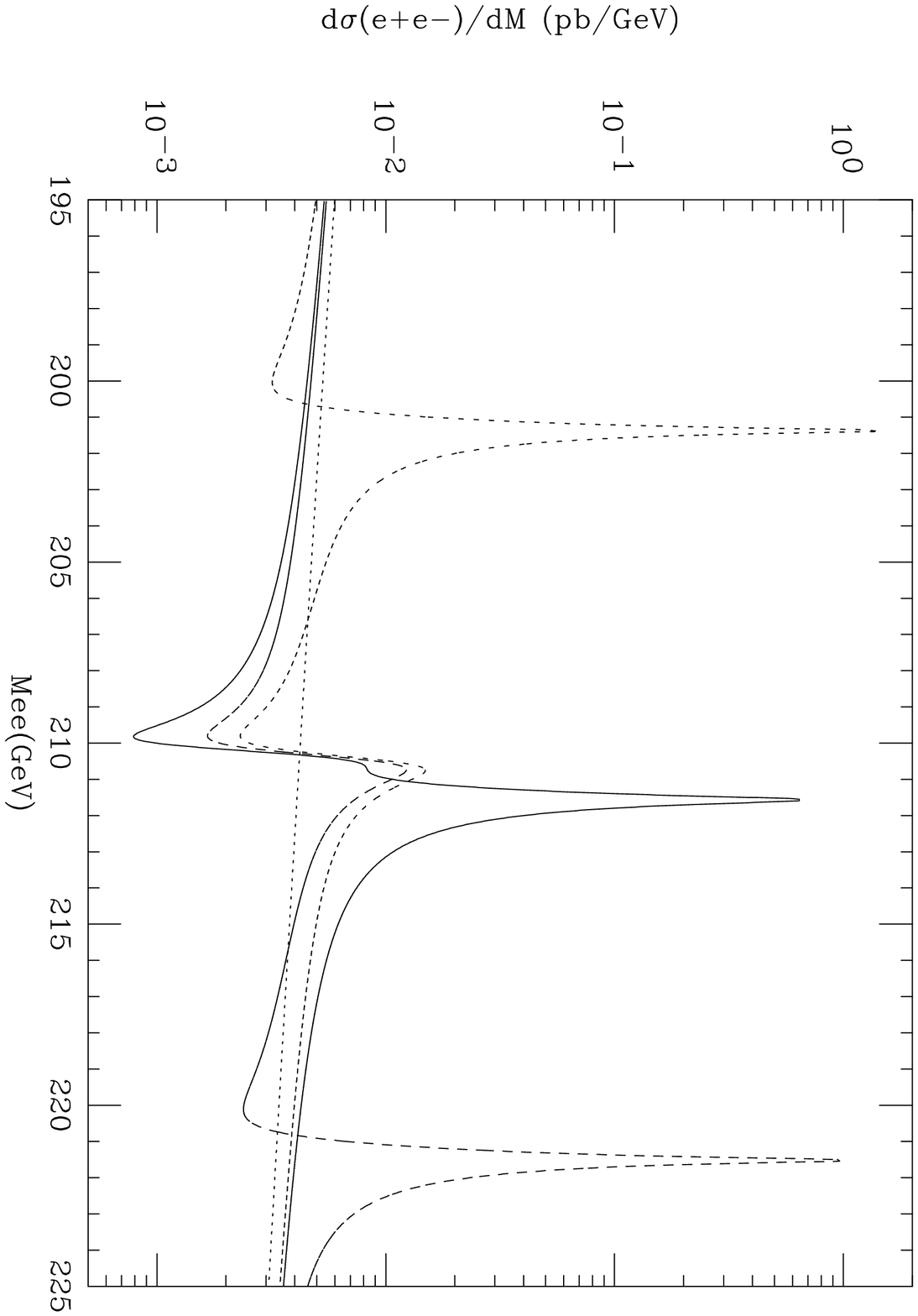}
\caption{Invariant mass distributions in $p \ol p$ collisions at $\ecm =
  2\,\tev$ for $\tom$, $\troz \ra e^+e^-$
  for $M_{\tro} = 210\,\gev$ and $M_{\tom} = 200$ (short-dashed curve), 210
  (solid), and $220\,\gev$ (long-dashed); $M_V = 500\,\gev$. The standard
  model background is the sloping dotted line. $Q_U + Q_D = \frac{5}{3}$ and
  $M_{\tpi} = 110\,\gev$; from Ref.~\cite{Lane:1999uh}.
  \label{TeV4LHC_fig_7}}
\end{figure}

So, to sum up, there are nagging little hints of something at $\sim
110\,\gev$ in dijets with a $b$-tag coming from some parent at $\sim
210\,\gev$. They've been around since Run~1 and it's time now to close the
book on them. I urge my experimental colleagues to settle this soon.

\section*{Acknowledgments}
I am grateful for many conversations with colleagues and friends in CDF and
D\O. I again thank Steve Mrenna for putting the TCSM into {\textsc{Pythia}}.
Thanks also to David Rainwater and Bogdan Dobrescu of the TeV4LHC Landscapes
project for their patience with me. This research was supported in part by
the Department of Energy under Grant~No.~DE--FG02--91ER40676.

\vfil\eject

\bibliography{TeV4LHC}
\bibliographystyle{utcaps}
\end{document}